\documentclass[%
reprint,
 amsmath,amssymb,
 aps,
]{revtex4-1}

\usepackage{physics}
\usepackage{lineno}

\usepackage[dvipdfmx]{graphicx}
\usepackage{bm}
\usepackage{color}

\begin{document}

\preprint{APS/123-QED}

\title{Room-temperature addressing of single rare-earth atoms in optical fiber}

\author
{Mikio Takezawa$^{1}$, Ryota Suzuki$^{1}$, Junichi Takahashi$^{1}$, Kaito Shimizu$^{1}$, Ayumu Naruki$^{1}$, Kazutaka Katsumata$^{1}$, Kae Nemoto$^{2}$, Mark Sadgrove$^{1}$, Kaoru Sanaka$^{1\ast}$ \\
\normalsize{$^{1}$Department of Physics, Tokyo University of Science,}\\
\normalsize{Kagurazaka 1-3, Shinjuku, Tokyo 162-8601, Japan}\\
\normalsize{$^{2}$Okinawa Institute of Science and Technology Graduate University, 1919-1 Tancha,}\\
\normalsize{Onna-son, Kunigami-gun, Okinawa 904-0495, Japan}\\
}

\begin{abstract}
Rare-earth atoms in solid-state materials are attractive components for photonic quantum information systems because of their coherence properties even in high-temperature environments. We experimentally perform the single-site optical spectroscopy and optical addressing of a single rare-earth atom in an amorphous silica optical fiber at room temperature. The single-site optical spectroscopy of the tapered rare-earth doped fiber shows nonresonant emission lines similar to those seen in the case of an unstructured fiber, and the autocorrelation function of photons emitted from the fiber shows the antibunching effect due to the spatial isolation given by the tapered fiber structure. The ability to address single rare-earth atoms at room temperature provides a very stable and cost-effective technical platform for the realization of a solid-state system for a large-scale quantum optical network and other quantum technologies based on a large number of spectral channels from visible to mid-infrared wavelengths. 
\end{abstract}

\pacs{Valid PACS appear here}
\maketitle


\section{Introduction}\label{sec1}


Quantum information processing on fiber networks is expected to realize secure communications over long distances and distributed quantum computation \cite{kimble08, obrien09}.  In a typical model of such quantum networks, their nodes consist of long-lived quantum bits (qubits) and single-photon emitters.  Single photons from these emitters quantum mechanically connect qubits between nodes, enabling us to utilize the networks as one quantum information system \cite{aharanovich16,childress05}. The nodes are typically fabricated with nanostructures or nanoresonators that contain solid-state-matter qubits interfaced with photonic waveguides. 

Rare-earth (RE) atoms and ions in solid-state materials are promising candidates for a large variety of photonic quantum devices. They have good compatibility with fiber networks, and their wide spectral window from visible to midinfrared wavelengths would be ideal for a quantum information platform \cite{kinos21}. Fig.1a shows an overview of the applications of quantum information technology depending on the rare-earth emission wavelength \cite{valerii84}. The wavelengths from visible to near infrared, a range for which cost-effective single-photon detectors are available, are suitable for quantum logic operations as well as quantum random number generators \cite{luo20,oberreiter16}, sub-diffraction-limited superresolution imaging \cite{schwartz13,thiel09}, and scalable quantum computation \cite{knill01}. Quantum memory is essential for quantum communications and can also be demonstrated in the visible-to-infrared range;  it has recently been demonstrated using isotopes with a nonzero nuclear spin \cite{stegger12,maure12}.  By contrast, a wavelength of around 1.5 $\mu$m is the most suitable for fiber-based telecommunications, minimizing the photon loss in silica-optical-fiber cables \cite{aharanovich16}. The midinfrared wavelengths longer than 2.0 $\mu$m are  suitable for free-space telecommunications \cite{su18}  and imaging technologies for environmental sensing and biometrics \cite{weibring03,petersen14}.

\begin{figure}[h]%
\centering
\includegraphics[width=0.5\textwidth]{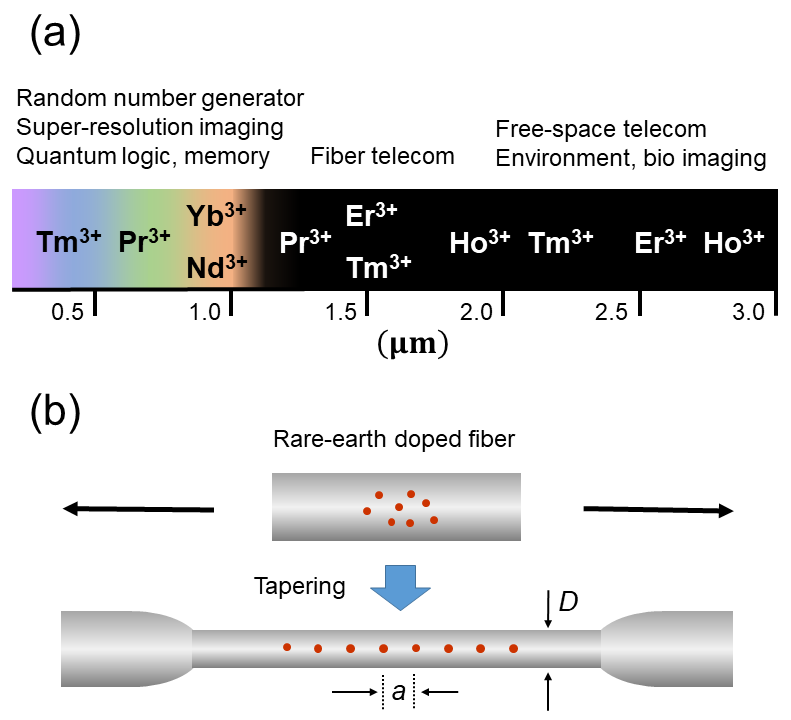}
\caption{ (a) An overview of the applications as quantum information technology depending on the rare-earth atom emission wavelength. (b) The construction process for tapered rare-earth doped optical fiber.  $D$ and $a$ are, respectively, diameter of the tapered fiber and the average distance between rare-earth atoms distributed in the fiber. }\label{fig1}
\end{figure}


Relevant to the several quantum applications listed above, the optical addressing of single RE atoms has been demonstrated using RE-doped solid-state materials, occasionally together with fabricated micro- or nanocavity structures. Examples include praseodymium-doped yttrium aluminum garnet  (Pr$^{3+}$:YAG) \cite{kolesov12}, yttrium orthosilicate (Pr$^{3+}$:YSO) \cite{utikal14}, and neodymium-doped yttrium orthovanadate (Nd$^{3+}$:YVO$_4$) \cite{zhong18}.  Optical addressing has also been demonstrated with fiber-based telecommunication wavelengths using erbium-doped silicon  (Er$^{3+}$:Si) \cite{yin13}, YSO (Er$^{3+}$:YSO) \cite{dibos18}, and titanium dioxide (Er$^{3+}$:TiO$_2$) \cite{dibos22}. The above demonstrations have typically been achieved using RE-doped crystalline materials at cryogenic temperatures. Here, we report an experimental demonstration of single-site optical spectroscopy and optical addressing of single RE atoms doped in an amorphous silica optical fiber at room temperature.

\section{Results}\label{sec2}

\subsection{Experimental setup}\label{subsec2}


We chose the ytterbium ion (Yb$^{3+}$) as the RE dopant for the fiber because of its attractive properties, as follows. First, the Yb$^{3+}$ ion has a simple energy-level structure with only two manifolds, $^2F_{7/2}$ (ground) and $^2F_{5/2}$ (excited), separated by the energy of approximately1.2 eV. This energy corresponds to the near-infrared region as shown in Fig. 1a, where laser diodes are readily available. Second, the excited state decays mainly radiatively with a long fluorescence lifetime of approximately 1 ms with high coherence. This characteristically quite long lifetime of RE atoms provides an environment for efficient population inversion and a technology base for fiber lasers and fiber amplifiers \cite{valerii84}. In contrast, however, the long lifetime also consequently leads to a low brightness for individual ions due to the Fourier-transform relation. This low photon emission rate is expected to be improved through the introduction of cavity structures, which allows us to utilize RE ions as a platform for quantum-information technologies.

Moreover, there are two naturally abundant isotopes with a non-zero nuclear spin: $\rm ^{171}  Yb^{3+}\ (I=1/2)$ and $\rm ^{173}  Yb^{3+} \ (I=5/2)$, where hyperfine transitions could be used as qubits in quantum memory applications \cite{welinski16}. A recent experimental work performed spin initialization, coherent optical and spin manipulation, and high-fidelity single-shot optical readout of the hyperfine spin state of $\rm ^{171}  Yb^{3+}\ (I=1/2)$ in a crystalline material \cite{kindem20}. These results strongly exhibit the usefulness of Yb$^{3+}$ ions using long spin coherence for realizing future quantum networks.


Fig. 1b shows the fabrication process for a tapered RE-doped optical fiber. The device was fabricated by tapering a commercially available ytterbium-doped fiber (nLIGHT, Inc.,  YB1200-4/125) using a heat-and-pull technique with a programmable stepping motor system and the methodology of the fabrication procedure is similar to that used in previous works \cite{uemura21,ward06}. The typical doping level of Yb$^{3+}$  into the fiber is $10^3 \sim10^4$ ppm by weight of Yb \cite{paschotta97}. This doping level corresponds to an ion density $10^{25} \sim \rm 10^{26} \ m^{-3}$. In the unstructured fiber, with $4-\mu$m diameter, the mean separation between Yb$^{3+}$  ions is estimated to be $10^{-14} \sim 10^{-13}$ m by simply calculating from the doping level.  The intensity and transversal modes of the excitation laser determine the actual number of optically active ions.

Several samples were  fabricated to study the average distance $a$ between single RE atoms distributed in the fiber, depending on the fabricated fiber diameter $D$. Fig. 2 shows a schematic of our experimental setup for the microscopy, spectroscopy, and autocorrelation measurement of single RE atoms in the tapered fiber. The collection efficiency for photons emitted from the atom inside the fiber into the guided modes is expected to reach about $32\%$ with our setup (see Appendix A). However, it will be necessary to excite the single atom selectively from the outside of the fiber for the realization of guiding the photons into the fiber. Observation of photons collected from the side of the tapered fiber is more technically feasible than the selective excitation because spatial-mode filtering is available, as shown in Fig. 2.

\begin{figure}[h]%
\centering
\includegraphics[width=0.5\textwidth]{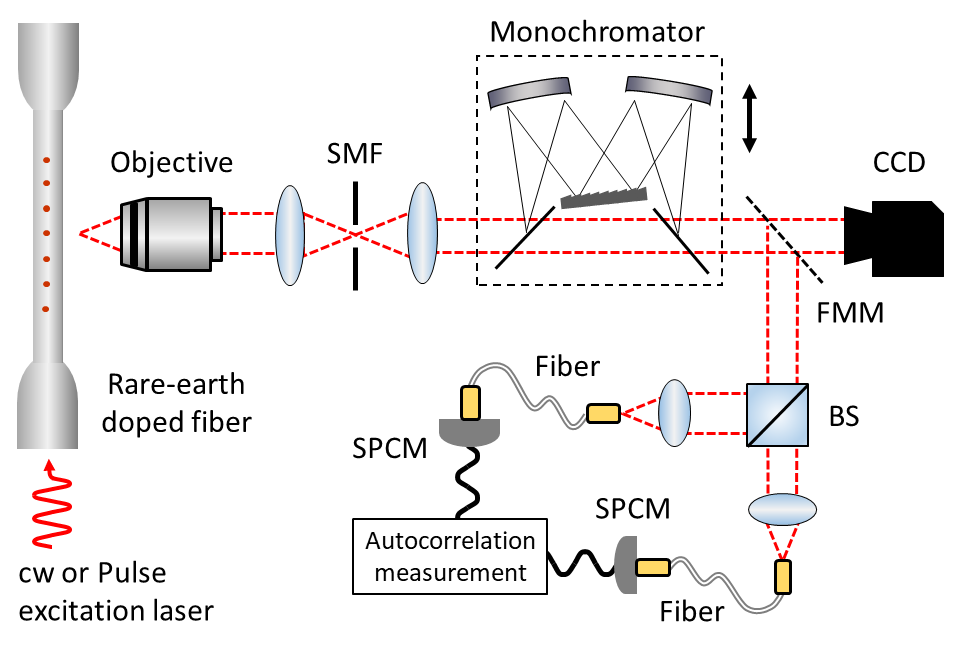}
\caption{A schematic of our experimental setup for the measurement of single rare-earth atoms in tapered fiber. A tapered fiber sample is placed on a metal stage at room temperature. A cw (wavelength 909 nm) or pulse (wavelength 980 nm wavelength, width 6 ns, and repetirion rate 10 MHz) excitation laser is injected into the tapered fiber input as a pump beam. The emitted photons are collected by an objective lens (40x magnification, N.A=0.65) and sent to the setup of the spatial-mode filter (SMF) composed of a pinhole and two lenses. A monochromator (SPG-120IR, Shimazu) is placed on a movable stage for the spectroscopy of photons emitted from the fiber. A CCD camera is used to obtain image pictures and the average distance between optically active rare-earth atoms in the tapered fiber. A flip-mirror mount (FMM) sends the photons into the setup of the second-order autocorrelation function. The input photons are separated using a nonpolarizing beam splitter (BS) and collected by multimode fibers and single-photon counting modules (SPCMs) with a typical detection efficiency $20-30\%$ for the observed wavelength range.}\label{fig2}
\end{figure}

\clearpage

\subsection{Microscopy}\label{subsec2}


\begin{figure}[h]%
\centering
\includegraphics[width=0.5\textwidth]{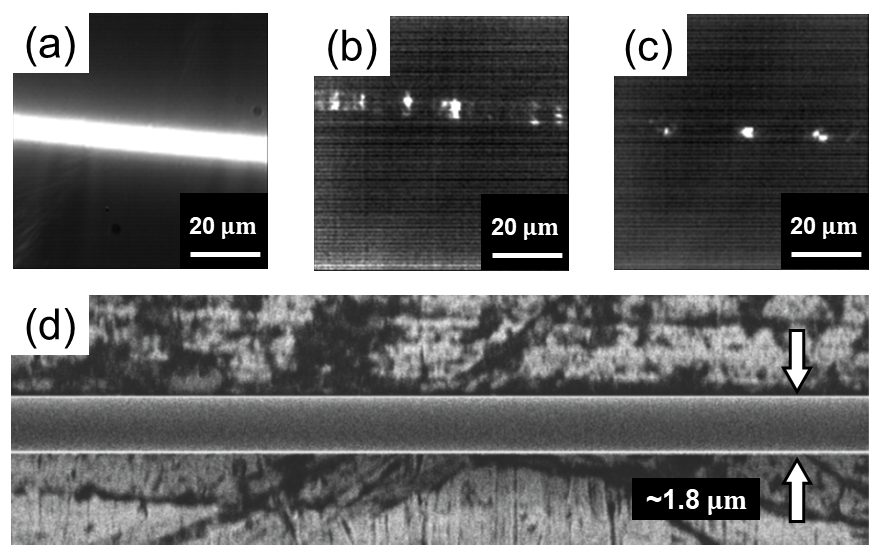}
\caption{Images of unstructured and tapered fibers taken by CCD camera with diameters of a (a) $\rm 4.4 \ \mu m$ , (b) $\rm 2.0 \ \mu m$, and (c)  $\rm 1.8 \ \mu m$. (d) Typical SEM image of structured tapered fiber with a diameter of $\rm 1.8 \ \mu m$  around its center. }\label{fig3}
\end{figure}

Fig. 3a-c show images taken by a CCD camera. Fig. 3d shows that taken by a scanning-electron-microscope (SEM) image. The scale bars in Fig. 3a-c have been measured by the micrometer position of the fiber sample stage. The scale of the fiber diameters shown in Fig. 3d has been measured directly by SEM. Note that the fiber and spot sizes seen in Figs. 3a-c appear larger than their real size due to several factors, including the light haze effect due to insufficient focusing conditions and the resolution limit of the optical setup. In the unstructured fiber ($D$ = 4.4 $\mu$m), photons from the RE atoms are emitted from every part of the fiber as shown in Fig. 3a. In contrast, photons from the tapered fiber ($D$ = 2.0, 1.8 $\mu$m) are emitted from discrete spots as shown in Figs. 3b and 3c. It appears that the average distance $a$ between distributed single RE atoms in the tapered fiber becomes much larger than the optical diffraction limit determined by the wavelength of the emitted photons, which is approximately 1 $\mu$m. In fact, from the pictures only, it is difficult to distinguish whether the separated spots are due to a single ion or whether the spots include multiple ions. A single ion tends to show a sharp, small-area spot as shown in Fig. 3c. The determination of whether or not the spot is produced by a single ion is actually made after the fact by performing an autocorrelation measurement. The observation of these spots becomes extremely difficult in regions of the tapered fiber with smaller diameters. The appropriate diameter to isolate individual RE atoms is estimated to be about 2 $\mu$m.

The various diameters and the measured average distances are summarized in Table I. The relation between the diameter and the average distances is not simply determined by the ion density of Yb$^{3+}$ and the structure sizes.  The condition of the transverse guided mode of the fiber becomes drastically different from unstructured to tapered fibers. A simple calculation shows that the spatial distribution of the guiding mode of tapered fibers is composed of multiple transverse modes, because the fiber core is no longer surrounded by a cladding layer after the heat-and-pull procedure but directly exposed to the air. The guided mode of the fiber is distributed mainly around the circumference of the optical fiber instead of the center due to the multiple transverse modes. This result indicates that the Yb$^{3+}$ ions, which are found at the fiber center, are excited relatively inefficiently by the guided excitation light. In principle, it is possible to isolate single ions with larger diameters if the doping level is sufficiently low. However, it is practically difficult to optically address single ions in an unstructured fiber because the ions become optically active more easily when the transverse mode of the fiber is near to the single-mode condition.

\begin{table}
\caption{The fiber diameters $D$ and the measured average distances of spots $a$}
\begin{center}
\begin{tabular}{|l|c|}
\hline
\ \ \ \ \ $D$ \ & Measurement results \\ \hline
\ \ 4.4 $\mu$m \ &\ \ Too dense to measure $a$ \ \ \\
\ \ 2.0 $\mu$m \ &\ $a$=23$\pm$3 $\mu$m \ \\
\ \ 1.8 $\mu$m \ &\ $a$=30$\pm$5 $\mu$m  \ \\
\ \ 1.3 $\mu$m \ &\ Scarcely observable \ \\
\ \ 1.0 $\mu$m \ &\ Not observable \ \\
\hline
\end{tabular}
\end{center}
\end{table}


\subsection{Spectroscopy}\label{subsec2}


\begin{figure}[h]%
\centering
\includegraphics[width=0.5\textwidth]{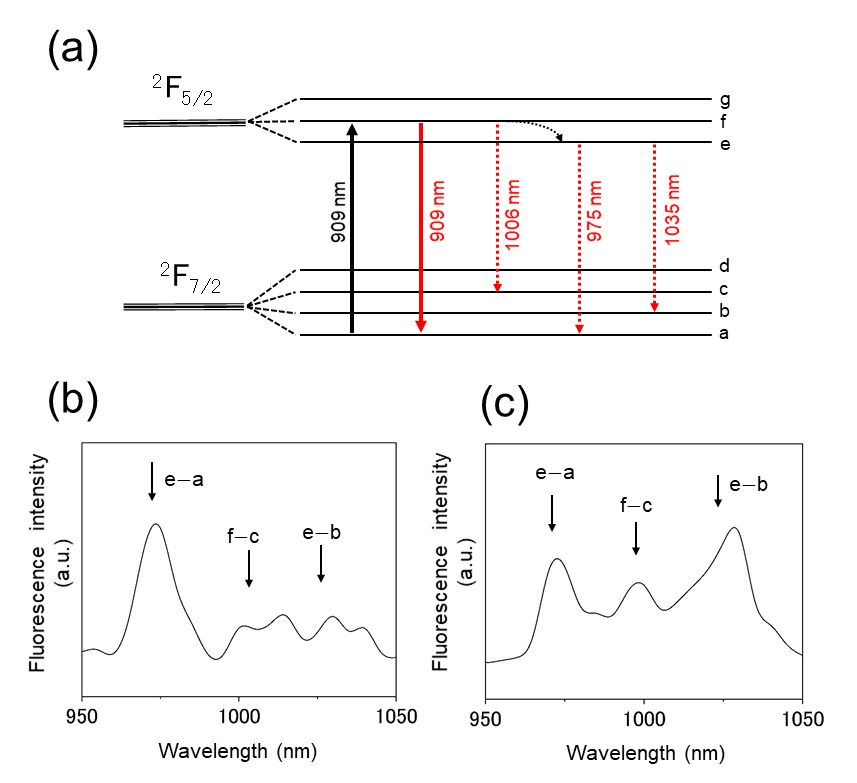}
\caption{ (a) The energy-level structure of ytterbium ion (Yb$^{3+}$) in silica glass pumped with a cw-excitation laser of wavelength 909nm. The emissions are assumed to be either the resonant fluorescence of wavelength 909 nm (solid line) or nonresonant fluorescence (dotted line). (b) The emission spectrum of photons emitted from unstructured Yb-doped fiber ($D$=4.4 $\mu$m). (c) The single-site emission spectrum of photons emitted from tapered Yb-doped fiber  ($D$=1.8 $\mu$m).}\label{fig4}
\end{figure}

Fig. 4a shows the energy-level structure of the Yb$^{3+}$ ion in silica glass. It also shows one pumping line with a wavelength of 909-nm and four dominant emission lines. The 909nm emission line is the resonant fluorescence, whereas the other three lines are nonresonant fluorescences that are observable within our spectroscopy measurement range. Fig. 4b shows the measured spectrum of photons emitted from an unstructured Yb-doped fiber ($D$=4.4 $\mu$m).  The emission of the resonant transition is not shown because the intensity of the resonant transition is much higher than the nonresonant transitions under the same excitation power. This spectrum has been investigated in detail for development of fiber lasers and fiber amplifiers \cite{valerii84,pask95,pas97}. There are more emission peaks in the spectrum than the assigned transition given by the energy level shown in Fig. 4a. The reason for  the additional peaks is likely due to the positions of Yb$^{3+}$ ions in unstructured fiber. The energy level given by Yb$^{3+}$ ions can be described by a simple two-level state. The sublevels of the ground and excited states would be energy degenerate in vacuum, but that degeneracy is removed by the electric field of the optical-fiber mode. Therefore, the energy sublevels would be different due to the difference of the electric fields, which depends on the position of the ions in the fiber.

Fig. 4c shows the single-site spectrum of photons emitted from a tapered Yb-doped fiber  ($D$=1.8 $\mu$m) using a cw-excitation laser (wavelength 909 nm, approximately 10 $\mu$W). The emitted fluorescence is assumed to be obtained in a steady state where the excitation and radiation rates match. Although the intensity ratios of the emission peaks are different, the three nonresonant fluorescence lines seen in the case of the unstructured fiber are still dominant even after tapering the fiber. No clear difference between the line width of the peaks can be distinguished between the unstructured and tapered fibers within the resolution limit of the monochromator, of approximately 2 nm. The intensity ratios of the spectrum from site to site are slightly different depending on the position of the emitter in the fiber. The reason for the difference of the intensity ratios is likely attributable to multiple factors related to the cylindrical structure of the tapered fiber itself. The transmission efficiency of photons from the inside of the fiber is affected by the position of the emitters due to the existence of the region of total internal reflection. Furthermore, the resonance condition depends on the wavelength of the emitted photons and the precise fiber diameter. (see Appendix B).

\begin{figure}[h]%
\centering
\includegraphics[width=0.5 \textwidth]{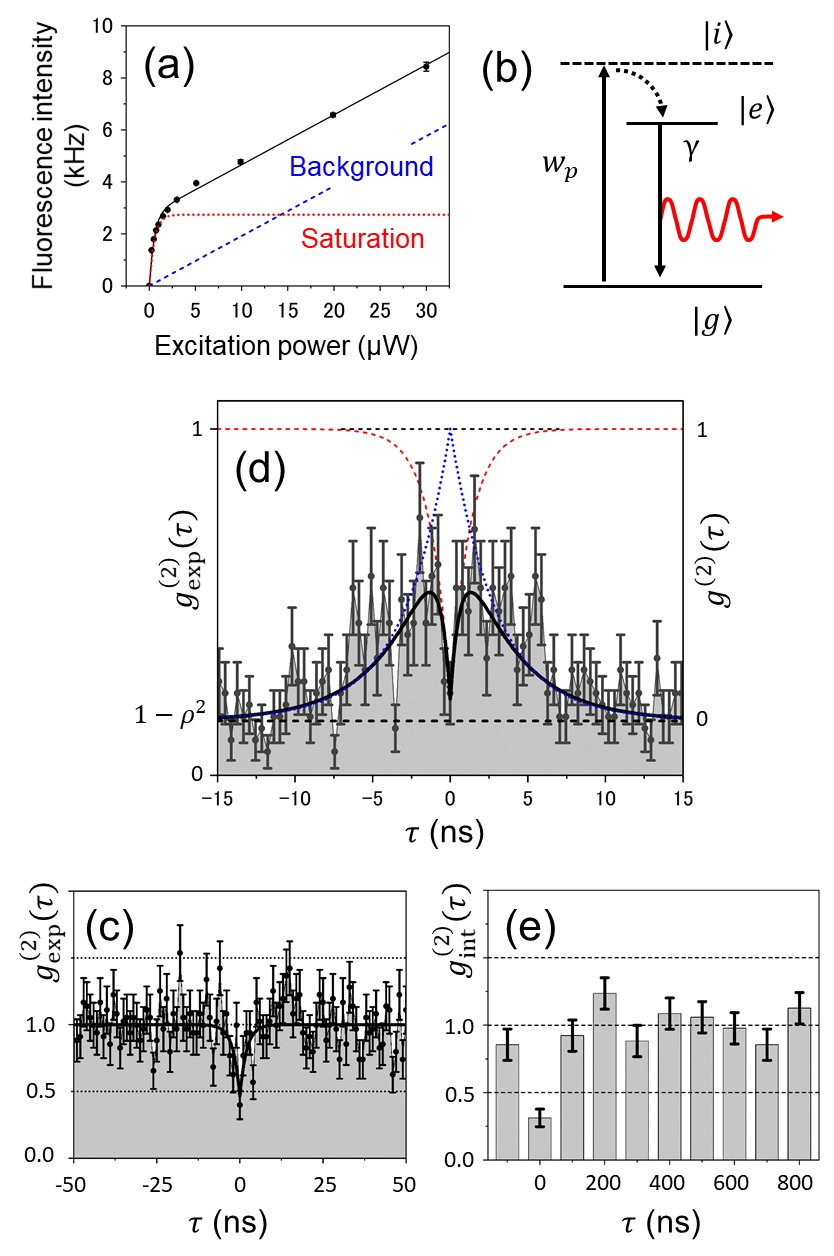}
\caption{(a) The fluorescence intensity as a function of the excitation laser power from a single Yb$^{3+}$ ion in tapered fiber (black dots) and a fitting curve (black solid line). The actual fluorescence intensity from a single Yb$^{3+}$ ion (red dotted line) is estimated by subtracting the linearly increasing background intensity (blue dashed line).  (b) A graphical illustration of a three-level system with excited states $\ket{i}$, and $\ket{e}$, and a ground state $\ket{g}$, where $w_p$ is the effective pump rate from $\ket{g}$ into $\ket{e}$ and $\gamma$ is the spontaneous radiative decay rate from $\ket{e}$ into $\ket{g}$. (c) The coincidence-measurement results with a fitting curve at zero-delay time obtained using a cw laser (wavelength 909 nm). The dots show the normalized coincidence counts with estimated errors. (d) The coincidence-measurement results with a fitting curve at zero-delay time obtained using a pulse laser (wavelength 980 nm, width 6 ns, and repetition rate 10 MHz). The dots show the normalized coincidence counts with estimated errors. The blue dotted line shows the envelope function given by the pulse laser width of 6ns. The red dashed line shows the estimated second-order autocorrelation function without the envelope function. (e) The second-order autocorrelation functions from the relationship of the coincidence counts measured in the repetition rate of the pulse laser with no uncorrelated backgrounds.}\label{fig5}
\end{figure}

\clearpage

\subsection{Autocorrelation}\label{subsec2}

Fig. 5a shows the intensity of fluorescence emitted from a single Yb$^{3+}$ ion in the tapered fiber ($\rm D=1.8 \ \mu m$) as a function of the excitation laser power. By considering the linearly increasing background intensity, we can estimate the excitation power to saturate a single Yb$^{3+}$ ion as $\rm 0.54\pm0.07 \ \mu W$ from the fitting curve. Fig. 5b shows a graphical illustration of a three-level system with the intermediate excited state $\ket{i}$, the radiatively active upper state $\ket{e}$, and the ground state $\ket{g}$, which is the simplest model for the transition process of photons emitted from a single atom. Here the ground state corresponds to the state $a$ in Fig. 4a.  The intermediate excited state and the radiatively active upper state are regarded as almost energy degenerate and correspond to the state $f$  in Fig. 4a.

We assume that the transition from $\ket{i}$ into $\ket{e}$ is much faster than the other transitions, where $w_p$ is the effective pump rate from $\ket{g}$ into $\ket{e}$, and $\gamma$ is the spontaneous radiative decay rate from $\ket{e}$ into $\ket{g}$. Theoretically, the second-order autocorrelation function $g^{(2)} (\tau)$ is obtained by solving the rate equation $\dot{\rho}_e=w_p \rho_g -\gamma \rho_e$ in the model of a three-level system \cite{michler00}, where $\rho_e (\tau)$ and $\rho_g (\tau)$ are the populations of energy levels $\ket{e}$ and $\ket{g}$, respectively. The function $\rho_e (\tau)$ is determined by solving the equation under the condition $\rho_g  (\tau)+ \rho_e  (\tau)=1$ as
\begin{align}
\rho_e (\tau) =\frac{w_p}{w_p+\gamma} \ \left\{1- \ e^{-(w_p+\gamma) \tau} \right\} +\rho_e (0) e^{-(w_p+\gamma) \tau}, \label{rho}
\end{align} 
where $\rho_e (0)$ determines the irremovable background coincidence counts at zero delay in the actual time-correlation measurement. The second-order autocorrelation function is defined by normalizing $\rho_e (\tau) $ with $\rho_e(\infty)$, resulting in
\begin{equation}
\label{g2general}
g^{(2)} (\tau)=1-\left\{ 1-g^{(2)} (0) \right\} \ e^{-(w_p+\gamma) \tau} . 
\end{equation}
The theoretical autocorrelation function (\ref{g2general}) shows that the antibunching time width appearing in the coincidence measurements corresponds to $2/(w_p+\gamma)$. 

We have measured the second-order autocorrelation function $g^{(2)} (\tau)$ for a single Yb$^{3+}$ ion using the same excitation laser to acquire the data shown in Fig. 5a, under the condition of less than saturation. Due to the weak excitation power of the laser, the count rate of the fluorescence intensity is limited to 1-2 kHz. Fig. 5c shows the measurement results of the coincidence counts at a zero-delay time of around $\pm$50 ns. A dip appears at the zero delay of the coincidences with a time width of approximately 4 ns. The measurement time window of the coincidences, about 100 ns is about a 4 orders of magnitude smaller than the spontaneous radiative decay time, which is estimated as $\gamma^{-1} \sim $ 1 ms \cite{valerii84}. Under the measurement-time areas, $2/w_p$ practically determines the antibunching time width and the autocorrelation function given in (\ref{g2general}) can be represented as $g^{(2)} (\tau) \simeq 1-\{ 1-g^{(2)} (0) \} \ e^{-w_p \tau} $. We have numerically fitted the function to the experimental results and estimated the second-order autocorrelation function at the zero delay as  $g^{(2)} (0) = 0.46\pm0.10$ and the time width of the dip appearing at zero delay is estimated as $2/w_p = \rm 4.0\pm1.5$ ns.

The measured value of $g^{(2)} (0)$ is likely to be larger than the actual value due to the influence of uncorrelated background coincidence events caused by dark counts of the detection system or other light sources such as room light in the time-correlation measurements. Therefore, we have used a pulsed laser (wavelength 980 nm, 6 ns pulse width, and repetition rate10 MHz) as the pump beam to evaluate the uncorrelated background coincidences and the fluorescence given by the laser with a wavelength of 980 nm is likely under nearly resonant conditions. Fig. 5d shows the measurement results of the coincidence counts at a zero-delay time of around $\pm$15 ns. When we use a pulsed laser, the autocorrelation function given in (\ref{g2general}) is outlined by the envelope function of the pulse shape. Here, we suppose that the envelope function is represented by a simple exponential function that has time width $\tau_o$. Therefore, the autocorrelation function given in (\ref{g2general}) should be modified as follows:
\begin{equation}
\label{g2pulse}
g^{(2)} (\tau) \simeq e^{-\frac{2 \tau}{\tau_o}}  \left[\ 1-\left\{ 1-g^{(2)} (0) \right\} \ e^{-w_p \tau} \ \right] . 
\end{equation}

We consider the uncorrelated background coincidence events to improve the experimentally observed second-order autocorrelation function at zero delay. The effect of background with average intensity $I_{\rm bg}$ on $g^{(2)} (\tau)$ for an emitter of intensity $I_{\rm em}$ can be derived from the experimentally measured second-order autocorrelation function, defined as  $g_{\rm exp}^{(2)} (\tau) = 1-\rho^2+ \rho^2 g^{(2)} (\tau)$, where $\rho$ is the ratio of the emitter to the total intensity, defined as $\rho=I_{\rm em}/(I_{\rm em}+I_{\rm bg})$  \cite{brouri00,fishman23}. Referring to the autocorrelation function under pulsed excitation given in (\ref{g2pulse}), the experimentally measured second-order autocorrelation function can be finally represented as:
\begin{align}
 g_{\rm exp}^{(2)} & (\tau) \simeq 1-\rho^2 \nonumber \\
  & + \rho^2  e^{-\frac{2 \tau}{\tau_o}} \left[\ 1-\left\{ 1-g^{(2)} (0) \right\} \ e^{-w_p \tau} \ \right] . \label{g2exp}
\end{align} 

The relation between $g_{\rm exp}^{(2)}(\tau)$ and $g^{(2)}(\tau)$ is displayed on the left-hand and right-hand vertical axis in Fig. 5d. We have numerically fitted the above function to the experimental results as shown in Fig. 5d, with a fixed pulse width $\tau_o$ of 6 ns. The time width of the dip, which appears at zero delay, is estimated as $2/w_p = \rm 2.6 \pm 0.2 \ ns$. The ratio of the emitter to the total intensity is estimated as $\rho = \rm 0.92 \pm 0.02$. This result matches the typical single-photon count rates of 1-2 kHz as opposed to the directly measured dark-count rate of about 100-200 Hz. The experimentally measured second-order autocorrelation function at zero delay is estimated as  $g_{\rm exp}^{(2)} (0) = 0.2 \pm 0.1$. The second-order autocorrelation function without the uncorrelated background coincidence events is estimated as  $g^{(2)} (0) = 0.1 \pm 0.1$. Both $g_{\rm exp}^{(2)} (0)$ and $g^{(2)} (0)$ show values of less than 0.5 meaning that the observed photons are emitted from an isolated single Yb$^{3+}$ ion in the tapered fiber. 

Finally, the antibunching effect was observed with two different types of excitation laser, providing convincing evidence that the photons are emitted from a true single Yb$^{3+}$ ion. In principle, it is possible to increase the width of the zero-delay dip in Fig.5c and 5d by reducing the power of the excitation laser because the width is practically determined by $2/w_p$. However, it is not technically feasible to perform the measurements of the coincidence counts because measurement with single-photon count rates less than 1 kHz takes too long to accumulate the required data under stable conditions. Therefore, we have changed the measurement time window of the coincidences from 100 ns to 1 $\mu$s to obtain about a 10 times higher rate of coincidence counts, and we have measured the second-order correlation from the relationship of the coincidence counts measured in the repetition rate of the pulse laser as shown in Fig. 5e. On the other hand, the ratio of the emitter to the total intensity estimated from uncorrelated backgrounds deteriorates as $\rho = \rm 0.64 \pm 0.04$ due to the low single-count rates less of than 1 kHz. We have removed the estimated uncorrelated background coincidences from all peaks in the time window, binned the coincidence counts of the peaks over approximately 35 ns, and divided the peak at zero delay by the average of the side peaks. The experimentally measured second-order autocorrelation function at zero delay is $g_{\rm int}^{(2)} (0) = 0.31 \pm 0.07$, as shown in Fig. 5e. 

The value of $g_{\rm int}^{(2)} (0) $ theoretically corresponds to the integrated function shown in (\ref {g2pulse}) with normalization by the integrated envelope function of the pulse shape, and can be represented as follows:
\begin{align}
g_{\rm int}^{(2)} & (0) \simeq \frac{\int_{0}^{\infty} \ g^{(2)} (\tau) \ d \tau}{\int_{0}^{\infty} e^{-\frac{2 \tau}{\tau_o} } \ d \tau }  \nonumber \\
& =  1-\left (1+\frac{w_p \tau_o}{2}\right)^{-1} \ \left\{ 1-g^{(2)} (0) \right\}, \label{g2int}
\end{align} 
where we suppose that the integration time range is sufficiently larger  than the pulse width $\tau_o$.  The value of $2/w_p$ can be approximately represented by \\
$\tau_o \{1-g_{\rm int}^{(2)} (0)\}/\{g_{\rm int}^{(2)} (0)-g^{(2)} (0)\} $ from (\ref{g2int}). The estimated value of $2/w_p $ with low single-count rate of less than 1 kHz is 18$\pm$14 ns under the premise that $g^{(2)} (0) $ is similar to the value shown in Fig. 5d.

\section{Discussion}\label{sec3}

The saturation behavior shown in Fig. 5a also appears in other solid-state systems that emit single photons. Such emitters that work at room temperature include CdSe/ZnS nanocrystals on SiO$_2$ substrates \cite{michler00}, nitrogen-vacancy (NV) centers in diamond \cite{kurtsiefer00}, gallium nitride (GaN) quantum dots on SiO$_2$ substrates \cite{holmes14}, defects in GaN substrate \cite{berhane17}, and hexagonal boron nitride (h-BN) in cubic zirconia hemispheres \cite{zeng22}.  The above-mentioned single-photon emitters typically have saturation powers on the order of about hundreds of microwatts to a few milliwatts. In contrast, the saturation power for a single Yb$^{3+}$ ion in a tapered fiber, $\rm 0.5 \sim 1.0 \ \mu W$, is characteristically 2 or 3 orders of magnitude lower than those of conventional solid-state systems.  In the two-level energy model, the saturation power and the fluorescence lifetime are inversely related. Therefore, the millisecond-level fluorescence lifetime given by Yb$^{3+}$ ions offers the advantage of ease of saturation with low excitation power. Due to the easiness of saturation, the intensity attenuation of the excitation light is very small and the location dependence of the pump intensity can be almost negligible.

However, the ease of saturation seen here for a single Yb$^{3+}$ ion causes an unwanted background signal for the emitted photons and the very long radiative decay time on the order of milliseconds results in a low emission rate per unit time because of the relation between the emission rate and the saturation excitation laser power shown in Fig. 5a.  Although we have to suppress the excitation power for the emitter not to saturate a Yb$^{3+}$ ion, as shown in Fig. 5a, the emission rate of the photons needs to be higher than the dark-count rates of the detectors and the background counts. Under the very low saturation power, this makes it hard to satisfy both requirements simultaneously. 


Related to our method, optical-nanofibers system have also been attracting considerable attention in the field of quantum optics. The conventional methods are given by embedding single-photon emitters such as quantum dots into a polymer nanofiber externally \cite{gaio16} or combining the tapered fiber and externally embedded single-photon emitters from the outside of the fiber  \cite{fujiwara11,shafi20}. The channeling efficiency of the photons emitted from the externally embedded emitter into the fiber-guided mode is determined by the fiber-core sizes and the wavelength of the emitted photons. Under the condition of a small enough size of fiber cores that has about half of the wavelength (approximately 0.5 $\mu$m) , the maximum channeling efficiency of photons from the externally deposited emitter into the fiber-guided mode reaches a theoretical value of about 30$\%$  \cite{yalla12}.

In contrast, we have used initially doped rare-earth ions inside of the tapered silica fiber directly. The channeling efficiency of the photons emitted from the inside of the optical-fiber core into the fiber-guided mode is simply determined by the refractive index of the fiber $n$ for the emitted photons as $\eta =  1-1/n $  (see Appendix A). When the wavelength of theh photon is approximately 1 $\mu$m the channeling efficiency is about $32\%$, which corresponds to the maximum efficiency given by the method using externally deposited single-photon emitters. Our method realizes the optimal channeling efficiency with no dependence on the position of emitters or fiber-core sizes within the scope of geometrical optics. The easiness of processing of the emitter due to the use of silica glass substrates is also useful for adjusting the distribution density of ions and fabricating cavity structures on the fiber itself. Such a property may offer more physically stable platforms for fabricating micro- and nanocavity structures on the tapered fibers \cite{schell15}.

Although it is necessary to resolve the above technical issues and improve the quality and efficiency of  single-photon emission from Yb$^{3+}$, our method has various other advantages over existing methods. These include the overwhelmingly low manufacturing and processing costs of single-photon-emitting devices, the ability to produce various wavelengths from visible to midinfrared for photonic quantum applications as shown in Fig.1a, and the ability to operate without the need for cryogenic systems. From the results for the nanosecond-level antibunching time width shown in Fig. 5d and 5e, it is expected that the experimental value of $g^{(2)} (0)$ will be further improved by using excitation lasers with pico- or femtosecond pulse widths. Externally compatible micro- and nanocavity structures on the tapered-fiber device can also be implemented to improve the emission efficiency of single photons induced by the Purcell effect \cite{yalla14,tetsumoto15}. The Purcell effect due to cavities is also useful for the generation of single photons with a high degree of indistinguishability for quantum networking applications because the effect drastically improves the total decay rate of single-photon emission relative to the dephasing rate of the decay process \cite{pettit23}. State-of-the-art measurements have been performed for the realization of indistinguishable single-photon emission with RE ions with this strategy \cite{ourari23}.

Additionally, it is known that the confinement of the guided modes of the tapered fiber itself substantially affects the spontaneous emission process depending on the fiber-diameter sizes and the distance from the atom to the fiber \cite{kien05}. Photons emitted from the atoms in the tapered fiber are likely to have a shortened fluorescence lifetime due to a similar effect. Furthermore, the cylindrical structure of the tapered fiber itself is known to work as a cavity structure through whispering-gallery modes \cite{afshar14} and is expected to work as a cavity to reduce the fluorescence lifetime through the Purcell effect \cite{xia22}. The appropriate diameters to isolate individual RE atoms are estimated to be approximately 2 $\ \mu m$ as shown in the subsection 2.2. The typical emission wavelength from the Yb$^{3+}$ ion  is approximately 1 $\mu$m as shown in the subsection 2.3. Six different standing-wave whispering-gallery modes are possible for this diameter and wavelength due to the confinement effect of the fiber according to a simple geometric calculation (see Appendix B). Therefore, the photons emitted from the atom under these conditions may have shorter radiation times and better coherence than those from the atoms in other regions. It is expected that the introduction of these strategies will lead to significant improvements, reaching the level of practical use in the near future.

\section{Conclusions}\label{sec4}

We have experimentally performed the single-site optical spectroscopy and optical addressing of single RE atoms in a tapered silica optical fiber at room temperature. The optical spectroscopy of a single Yb$^{3+}$ ion in the tapered fiber shows nonresonant emission lines similar to those seen in the case of the unstructured fiber. The autocorrelation function at zero delay of photons emitted from a single Yb$^{3+}$ ion shows $g_{\rm exp}^{(2)} (0) = 0.2 \pm 0.1$ with uncorrelated background coincidence events, and $g^{(2)} (0) = 0.1 \pm 0.1$ without uncorrelated background coincidence events, which is a direct evidence of photon emission from a single Yb$^{3+}$ ion in the fiber. The time width of the autocorrelation function is phenomenologically explained by the model of a three-level system, and is determined in practice by the pump rate of the single Yb$^{3+}$ ion. For practical use, it is necessary to improve the quality and efficiency of our emitter.  We expect that the quality and efficiency will improve markedly with the application of recently developed techniques such as laser pulse controls and the use of micro- to nanocavity structures.  Our results hopefully provide a platform for developing a very stable and cost-effective technology for photonic quantum applications over a large number of spectral channels.

\section*{Acknowledgment}

Authors would like to thank Y. Arakawa, Y. Ota, S. Iwamoto, and H. Takayanagi for useful discussions. This work was supported by Tokyo University of Science Research Grant and JST Grant-in-Aid for Scientific Research (C) Grant No. 21K04931.

\clearpage

\appendix

\section{Channeling efficiency}

\begin{figure}[h]%
\centering
\includegraphics[width=0.4\textwidth]{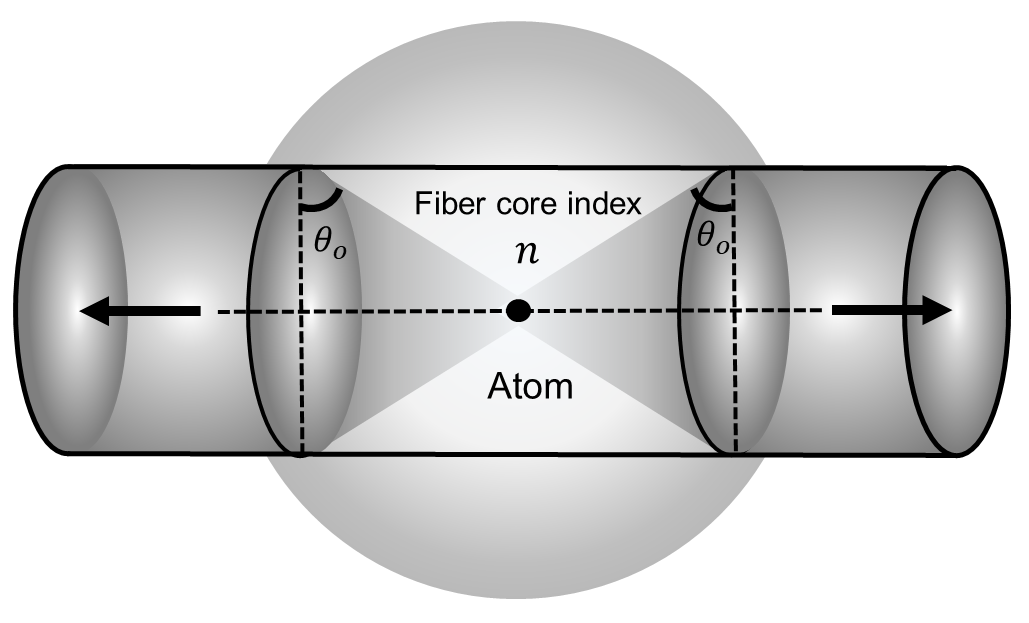}
\caption{Schematic illustration of an isolated single atom in a tapered fiber and the propagation of fluorescence photons emitted from the atom. Light shaded area shows the fluorescence by spontaneous emission. Dark shaded area shows the fluorescence into the guided modes of the fiber.}\label{figa}
\end{figure}


We have calculated the channeling efficiency of the fluorescence photons emitted from a single atom in a tapered fiber into the guided modes of the fiber. When an atom emitting single photons is placed in a tapered silica-optical-fiber core as shown in Fig. 6, the proportion of emitted photons that couple to the fiber-guided modes is determined by the critical angle $\theta_o$, which gives the total internal reflection inside the fiber. The critical angle $\theta_o$ is determined by the refractive index of the fiber core $n$ as $\sin \theta_o =1/n$.  From the geometrical relation of $\theta_o$, the solid angle $S$ from the atom to the fiber-guided mode is given by $S=2 \pi \{1-\cos(\pi/2-\theta_o) \}=2 \pi \{1-\sin \theta_o \}=2 \pi (1-1/n)$. 

The channeling efficiency of the fluorescence photons emitted from the single atom into the guided modes of the fiber, $\eta$, is given by dividing the above solid angle $S$ by the total solid angle $4 \pi$ as $\eta/2 = S/4 \pi = (1-1/n)/2 $. The factor $1/2$ for $\eta$ corresponds to the fact that we are considering fluorescence photons that couple to the rightward-propagating guided modes. Therefore the combined channeling efficiency from the single atom into both directions of the guided mode is given by $\eta =  1-1/n $.  The refractive index for a wavelength approximately 1 $\mu$m is $n \sim 1.45$. The combined channeling efficiency is about $32\%$ for this refractive index. The collection efficiency of photons from the atom into the guided modes is simply determined by the condition of total reflection; therefore, the coupling efficiency does not depend on the position of the atom and the size of the fiber-core within the scope of the geometrical optics. 

We also checked this value using finite-difference-time-domain (FDTD) simulations and found good agreement. When the emitter moves away from the fiber center, no large changes in the coupling are expected with a geometric coupling model. However, the emitter coupling can actually increase as well as decrease depending on its exact position due to the multimode nature of the fiber as follows. A dipole emitter at a point $\mathbf{r}_0$ with dipole moment $\mathbf{d}$ couples to a given mode with electric field profile $\mathbf{E}(\mathbf{r})$ with proportionality
\[C\propto \abs{ \mathbf{d}\cdot\mathbf{E}(\mathbf{r}_0) }^2.\] 
Finding the proportionality constant, or finding the relative coupling efficiency to the mode relative to all other modes
, is in general not trivial. In the case of vacuum-clad fibers, the analytical results are well known~\cite{kien05}. However, it is also possible to calculate the result numerically using FDTD or equivalent numerical methods. Additionally, in the limit where the fiber diameter $2a$ is much larger than the dipole-emitter wavelength $\lambda$, and the emitter is at the fiber center, ray-optics approximations may give sufficiently good results.

Here, we provide FDTD-calculated results for the coupling of a dipole emitter with a wavelength between 900 and 1040 nm, placed at the center of a vacuum-clad silica fiber, as shown in Fig. 7, and also for one displaced 500 nm from the center, as shown in Fig. 8. We assume that the dipole orientation is random and that the total coupling efficiency is therefore an average of the values for $x-$, $y-$, and $z-$oriented dipoles. The random-orientation assumption tends to be a good approximation for room-temperature quantum emitters~\cite{abe17}. For the sake of
interpreting the results, we will assume that the dominant modes to which the emitter can couple are the lowest-order modes HE$_{11}$ (the fundamental mode), TE$_{01}$, and TM$_{01}$ but we note that the FDTD simulations take into account coupling to all possible modes.

Note that if the emitter is at the fiber center, the $x-$ and $y-$dipole contributions are the same and the $z$-dipole contribution is relatively small, due to the vanishing $z$ component of the fundamental fiber mode at the fiber center. However, due to the fact that the present fiber is not in the single-mode regime for the considered wavelength range, nonvanishing coupling of the $z$ component is expected due to the presence of the TM$_{01}$ mode, which has a nonvanishing $z$ component at the fiber center. If the dipole is displaced from the center, the transverse component of the fundamental mode weakens in an approximately Gaussian fashion, while the longitudinal component increases. The TE$_{01}$ and TM$_{01}$ modes have a more complicated mode profile but generally increase in amplitude away from the fiber center, leading to increased coupling. Additionally, the contributions from $x-$ and $y-$oriented dipoles are no longer degenerate.

In the case currently under consideration, the fact that the rare earth ions are doped in the core of the fiber means that the ions are most likely to be near the fiber center. Nonetheless, some diffusion is possible during the heat-and-pull stage.  Therefore, for comparison, we simulate the case for a dipole at the center of the fiber and also for a dipole displaced 500 nm from the fiber center.

\begin{figure}[h]%
\centering
\includegraphics[width=0.4\textwidth]{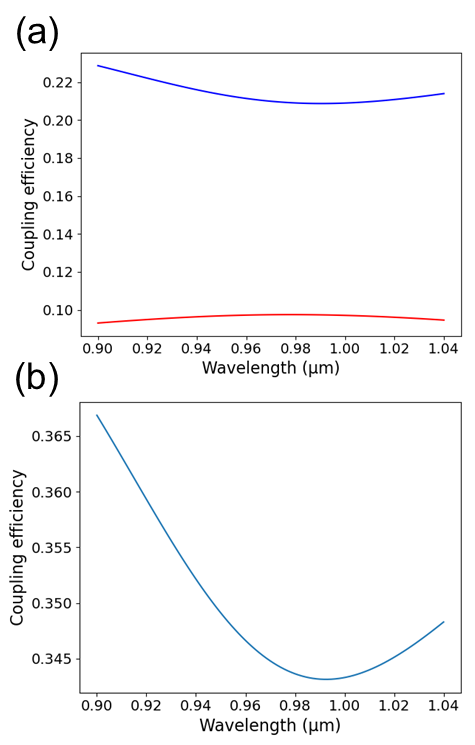}
\caption{The coupling efficiency for a dipole placed at the center of the fiber. (a) The contributions from transversely (blue) and longitudinally (red) oriented dipoles as indicated in the figure legend.  (b) The total coupling efficiency as a function of the wavelength.}
\label{figa2}
\end{figure}

\begin{figure}[h]%
\centering
\includegraphics[width=0.4\textwidth]{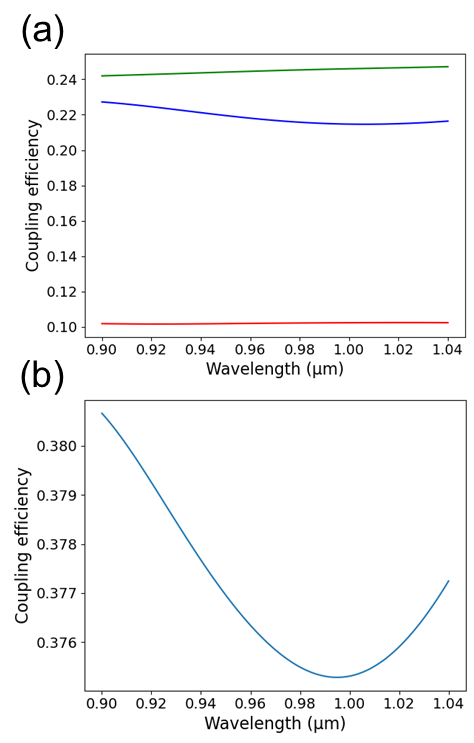}
\caption{The estimated coupling efficiency for a dipole offset 500 nm along the $y$ axis. (a) The contributions from $x-$(blue), $y-$(green), and $z-$(red) oriented dipoles as indicated in the figure legend. (b) The total coupling efficiency as a function of the wavelength.}
\label{figa3}
\end{figure}

Fig. 7a shows the results for a dipole placed at the fiber center. The main contribution to the coupled intensity comes from the transversely oriented dipole moment, with that for the longitudinal ($z$) component being relatively suppressed, despite the fact that the fiber fundamental mode is hybrid in nature and has a nonvanishing $z$ component in general.
However, this $z$ component is stronger near the surface of the fiber, due to the fiber boundary conditions, leading to the small contribution seen here.
Note that the total coupling at the representative wavelength of $1.0\;\mu$m is 34$\%$ in good agreement with a geometric coupling model as shown in Fig. 7b.

On the other hand, as seen in Fig. 8a, a dipole displaced from the center can counterintuitively have a larger coupling to the fiber, due in part to the increased contribution from the
$z-$oriented dipole moment and also due to the higher-order modes, which have larger intensities away from the fiber center. Here, the total coupling at the representative wavelength of $1.0\;\mu$m is increased to approximately 38 $\%$, as shown in Fig. 8b.


\section{Confinement efficiency}

\begin{figure}[h]%
\centering
\includegraphics[width=0.5\textwidth]{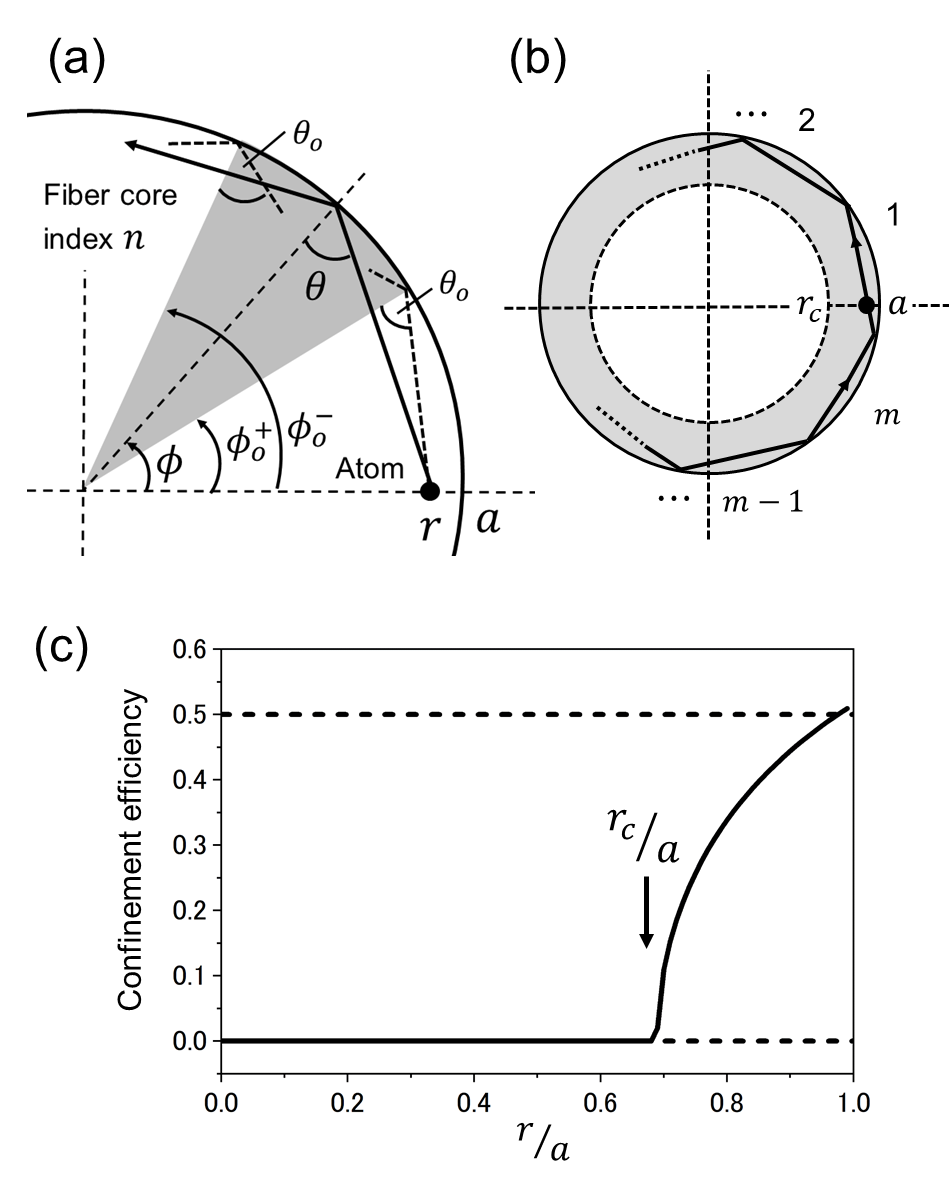}
\caption{(a) A schematic illustration of an isolated single atom placed in a tapered fiber with radius $a$.  The shaded area shows the azimuthal angle, which allows total internal reflection of photons emitted from the atom placed distance $r$ away from the fiber-core center. (b) A schematic illustration showing an isolated single atom under the standing-wave condition. The shaded area shows the regions of the confinement effect given by total internal reflection inside the fiber.  (c) A numerical evaluation of the confinement efficiency of fluorescence photons depending on the single-atom position in the radial direction with the wavelength $\lambda  \sim 1 \ \rm{\mu m}$ inside the silica glass fiber of radius $a \sim 1 \ {\rm \mu m}$. }\label{figb1}
\end{figure}

\begin{figure}[h]%
\centering
\includegraphics[width=0.5\textwidth]{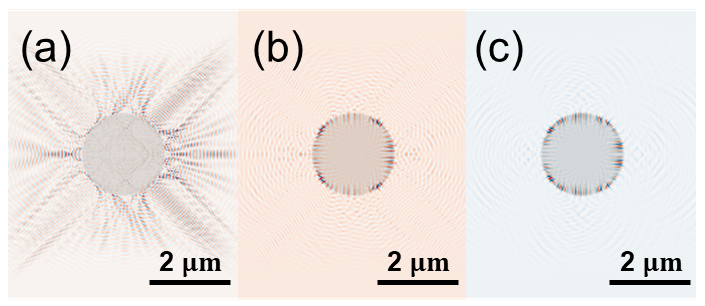}
\caption{The simulated transverse distribution of the emission intensity of a photon with the wavelength $\lambda  \sim 1 \ \rm{\mu m}$ inside a silica glass fiber of radius $a \sim 1 \ {\rm \mu m}$. The fiber core is shown by the shaded area. The atom positions from the center of the fiber $r$ satisfy  (a)  $r  < r_c$, (b) $r \sim r_c $, and (c) $r_c < r  <a$, where $r_c \sim a$ is the region in which total internal reflection exists. }
\label{fig:Center}
\end{figure}

We have calculated the confinement efficiency of fluorescence photons emitted from a single atom in a tapered fiber given by total internal reflection. We consider an atom emitting single photons placed a distance $r$ away from the fiber-core center, as shown in Fig. 9a. Here, the fiber-core radius is $a$ and satisfies $r<a$. The photons emitted from the atom are reflected with angle $\theta$ inside the fiber. The corresponding azimuthal angle to the reflection point is $\phi$. The angles and distances satisfy the following relational expression from their geometric positional relationship:
\begin{equation}
\label{angles}
r \sin \phi = \sqrt{a^2+r^2-2ar \cos \phi} \ \sin \theta. 
\end{equation}

The critical angle $\theta_o$ is determined by the reflective index of the fiber core, $n$, as $\sin \theta_o = 1/n$ and the condition for total reflection inside of the fiber core is given by $\sin \theta > 1/n$. When the reflection angle is $\theta_o$, the corresponding azimuthal angle could be calculated using \ref{angles} and two solutions $\phi_o^{+}$ and $\phi_o^{-}$ are given as follows:
\begin{equation}
\label{phi_o}
\phi_o^{\pm}=\cos^{-1} \frac{a}{n^2 r} \left[1 \pm \sqrt{1-n^2 \left\{ 1- \frac{(n^2-1) r^2}{a^2}  \right\} } \ \right]. 
\end{equation}

We obtain the inequality $r/a > 1/n$ from the condition for the azimuthal angles in \ref{phi_o} to be real-number solutions.  Therefore, the distance $r$ to allow total internal reflection inside the fiber core needs to satisfy  the inequality $r_c < r <a$ , where $r_c=a/n$ is the minimum distance to allow total internal reflection as shown by the shaded area in Fig. 9b. The ratio of the fiber area that allows total reflection is given by $p =\pi (a^2 - r_c^2) / \pi a^2=1-r_c^2/a^2$. When the wavelength of the emitted photon is approximately 1 $\mu$m, the refractive index of the silica fiber is $n \sim 1.45$. The minimum distance is $r_c \sim 0.68 a$ and the ratio is $p \sim 0.52$.  If we suppose that the atoms that emit single photons are randomly distributed in the fiber core, about half of them could emit photons that are confined inside of the fiber core. 

The range of the azimuthal angle that allows total internal reflection is given by $\abs{ \phi_o^{-}-\phi_o^{+} }$ as shown the shaded area in Fig. 9a. The angle area increases as the value of $r$ approaches $a$. We define the confinement efficiency of fluorescence photons as $\eta/2= \abs{ \phi_o^{-}-\phi_o^{+} }/2 \pi$.  The factor $1/2$ for $\eta$ corresponds to the fact that we are considering the fluorescence photons coupled only to the counterclockwise direction. Therefore, the combined confinement efficiency from the single atom into both the clockwise and counterclockwise directions is given by $\eta= \abs{ \phi_o^{-}-\phi_o^{+} }/ \pi$. Fig. 9c shows the confinement efficiency depending on the ratio of the distance from the core center $r/a$ with $n \sim 1.45$. The confinement efficiency approaches 0.5 as $r/a$ approaches 1. 

Standing waves are possible in the confinement region shown by the shaded area in Fig. 9b. The length of the half wavelength of the emitted photons inside the fiber is given by $\lambda/2 n$. The standing waves inside the region should satisfy the inequality $2\pi r_c < m \lambda/2n < 2\pi a$, where $m$ is the natural number of the half wavelength in the fiber. Here we consider the number of standing wave modes for the wavelength of approximately 1 $\mu$m in the fiber core with diameter approximately 2 $\mu$m. The numbers to satisfy the inequality are $m=13, 14, \cdots, 18$ with $\lambda = a \rm{\sim 1 \ \mu m}$ and $n \sim 1.45$. Therefore, the condition of the wavelength and the fiber size offers about six whispering-gallery modes, and the atom positions determine the coupling efficiency to those modes. 

We have also checked the value of the emission intensity using FDTD simulations under the above conditions and found good agreement. When the atom is placed at the position of  $r = 0.5a < r_c $, the photons emitted from the atom show hardly any confinement effect as, shown in Fig. 10a. When the atom is placed at the position of  $r =0.7a \sim r_c$,  whispering-gallery modes for the emitted photons start to appear around the surface of the fiber diameter as, shown in Fig. 10b. Finally, under this condition $r_c < r = 0.9a  <a$, the photon clearly resonates in this mode, as a standing wave as shown in Fig. 10c, and ray-optics approximations may give sufficiently good results.

\clearpage

\end{document}